\documentstyle[a4,12pt]{article}

\catcode`\@=11	

\def\@ddllist#1#2{\ifx#1\empty \else
 \ifx#2\empty \let#2#1\else \edef#2{#1,#2}\fi \fi}
\def\@ddrlist#1#2{\ifx#1\empty \else
 \ifx#2\empty \let#2#1\else \edef#2{#2,#1}\fi \fi}
\let\addlefttolist=\@ddllist
\let\addrighttolist=\@ddrlist
\def\@reorder#1{\def\@newlst{}\edef\@tmplst{#1}\def\@qm{?}%
 \@for\@nxtdum:=\@tmplst\do{\def\@bigbit{}%
 \@for\@nxtchr:=\@tmplst\do{\ifx\@bigbit\empty\def\@tmplst{}\fi
 \ifx\@nxtchr\@qm \@ddrlist\@nxtchr\@newlst \else \ifx\@bigbit\empty
 \let\@bigbit=\@nxtchr \else \ifnum\@nxtchr>\@bigbit
 \@ddrlist\@bigbit\@tmplst \let\@bigbit=\@nxtchr
 \else \@ddrlist\@nxtchr\@tmplst \fi \fi \fi}%
 \@ddllist\@bigbit\@newlst}\edef#1{\@newlst}}
\def\@collapse#1{\def\@newlst{}\def\@qm{?}\def\@tmpbit{}\def\@link{,}%
 \@for\@nxtchr:=#1\do{\ifx\@newlst\empty \let\@newlst=\@nxtchr \else
 \ifx\@nxtchr\@qm
 \edef\@newlst{\@newlst\@tmpbit,\@nxtchr}\def\@tmpbit{}\def\@link{,}\else
 \ifx\@lstchr\@qm \edef\@newlst{\@newlst,\@nxtchr}\else
 \ifnum\@tempcnta<\@nxtchr
 \edef\@newlst{\@newlst\@tmpbit,\@nxtchr}\def\@tmpbit{}\def\@link{,}\else
 \ifnum\@tempcnta=\@nxtchr
 \edef\@tmpbit{\@link\@nxtchr}\def\@link{--}\fi\fi\fi\fi\fi
 \ifx\@nxtchr\@qm \else \@tempcnta=\@nxtchr \advance\@tempcnta\@ne \fi
 \let\@lstchr=\@nxtchr}\edef\@newlst{\@newlst\@tmpbit}\edef#1{\@newlst}}
\let\collapselist=\@collapse \let\reorderlist=\@reorder
\def\citesort#1{\edef\@lst{#1}\@reorder{\@lst}\@collapse{\@lst}\edef#1{\@lst}}
\def\@citex[#1]#2{\if@filesw\immediate\write\@auxout{\string\citation{#2}}\fi
 \def\@citea{}\@for\@citeb:=#2\do
 {\ifx\@citea\empty\else \edef\@citea{\@citea ,}\fi
 \@ifundefined{b@\@citeb}{\edef\@citea{\@citea?}%
 \@warning{Citation `\@citeb' on page \thepage\space undefined}}%
 {\edef\@citea{\@citea\csname b@\@citeb\endcsname}}}%
 \@reorder\@citea \@collapse\@citea \@cite{\@citea}{#1}}
\def\abbstop{\futurelet\next\@@bbstp}			
\def\@@bbstp{\ifx\next.\else.\fi}			
\def\stopcomma{.\futurelet\next\@@bbscm}		
\def\@@bbscm{\ifx\next.\let\next\@@bbeat		
 \else\@@bbcma\let\next\relax\fi\next}
\def\@@bbeat#1{\futurelet\next\@@bbcma}			
\def\@@bbcma{\ifx\next,\else, \fi\ignorespaces}		
\catcode`\@=12     
\def\ital#1{{\it#1\/}}					
\def\mathintext#1{\ifmmode #1\else $#1$\fi}		
\def\open#1{I\mathchoice{\kern-0.90ex}{\kern-0.90ex}	
 {\kern-0.60ex}{\kern-0.45ex}#1}			

\def\vev#1{\left\langle{#1}\right\rangle}		
\def\unit{{\hbox to.34em{\kern.13em			
 \vrule height.73em width.05em\kern.07em\vrule height.79em width.05em
 \kern-.25em\hrulefill\kern-.36em\raise.41em\hbox{\char'40}\kern.06em}}}
\def\spig{\kern.3em\raise1.2ex\hbox{$|$}\kern-.48em\to}	
\def\slash#1{\rlap{\kern0.07em/}#1}			
\def\slashcap#1{\rlap{\kern0.25em/}#1}			
\def\hbar{{\mathintext{\mathchoice			
 {\rlap{\vrule height1.2ex width0.4em depth-1.1ex}}%
 {\rlap{\vrule height1.2ex width0.4em depth-1.1ex}}%
 {\rlap{\vrule height.84ex width.28em depth-.77ex}}%
 {\rlap{\vrule height.60ex width.20em depth-.55ex}}h}}}
\def\dddot#1{\vbox{\ialign{##\crcr			
 $\hfil .\>\!\!.\>\!\!. \hfil$\crcr\noalign{\kern1.7pt\nointerlineskip}
 $\hfil\displaystyle{#1}\hfil$\crcr}}}
\def\barcap#1{\vbox{\ialign{##\crcr			
 $\hfil \mkern3mu - \hfil$\crcr\noalign{\kern-2.3pt\nointerlineskip}
 $\hfil\displaystyle{#1}\hfil$\crcr}}}
\def\rectangleunits#1#2#3{{\vcenter{\vbox{		
 \hrule \hbox{\vrule height#2#3\kern#1#3\vrule}\hrule\vskip2pt}}}}

\def\triblockunits#1#2{{\vcenter{\vbox{\hbox{
 \dimen2=0pt\loop\ifdim#1#2>2\dimen2\relax\rlap{\raise1.75\dimen2\relax
 \hbox to#1#2{\kern\dimen2\hrulefill\kern\dimen2}}%
 \advance\dimen2by0.2pt\repeat\kern#1#2}\vskip2pt}}}}

\def\addendum{\ital{addendum}}		
\def\adhoc{\ital{ad hoc}}

\def\aposteriori{\ital{a posteriori}}	
		
\def\erratum{\ital{erratum}}

\def\naive{na{\"\i}ve}			
\def\etal{{\it et al}\abbstop}

\def\ibid{{\it ibid}\abbstop}		
\def\cf{{\it cf}\abbstop}		

\def\eg{{\it e}.{\it g}\stopcomma}	
\def\ie{{\it i}.{\it e}\stopcomma}
\def\units#1{\hbox{\,#1}}

\def\gev{\units{GeV}}

\def\gsim{\mathbin{\hbox{\rlap				
  {\raise0.4ex\hbox{$>$}}\lower0.7ex\hbox{\small$\sim$}}}}
\def\lsim{\mathbin{\hbox{\rlap				
  {\raise0.4ex\hbox{$<$}}\lower0.7ex\hbox{\small$\sim$}}}}
\def\JNO#1,#2,#3{{\bf#1}\ (19#2)~#3}
\def\ERR#1,#2,#3 {\erratum\ \JNO#1,#2,#3}
\def\ERRATUM#1,#2,#3 {\erratum\ \JNO#1,#2,#3}
\def\ADD#1,#2,#3 {\addendum\ \JNO#1,#2,#3}
\def\ADDENDUM#1,#2,#3 {\addendum\ \JNO#1,#2,#3}
\def\journal#1#2,#3,#4 {\ital{#1} \JNO#2,#3,#4\ }

\def\samejournal#1,#2,#3 {\JNO#1,#2,#3\ }
\def\IBIDEM#1,#2,#3 {\ibid\ \JNO#1,#2,#3\ }
\def\addjour#1,#2,#3 {\JNO#1,#2,#3\ }
\def\accjour#1#2{accepted for publication in \ital{#1}\if.#2\else~{\bf#2}\fi}
\def\appjour#1#2{to appear in \ital{#1}\if.#2\else~{\bf#2}\fi}
\def\subjour#1#2{submitted to \ital{#1}\if.#2\else~{\bf#2}\fi}

\def\JPG   {J.\ of Phys.\ G}

\def\NP    {Nucl.\ Phys\abbstop}

\def\PL    {Phys.\ Lett\abbstop}
\def\PR    {Phys.\ Rev\abbstop}
\def\PRL   {\PR\ Lett\abbstop}

\def\ZP    {Z.\ Phys\abbstop}
\def\Z{\hphantom{-}}	\def\Y{\hphantom{|}}
\def\gA{g_{\scriptscriptstyle A}} 	\def\gV{g_{\scriptscriptstyle V}}
 	
\def\mA{m_{\scriptscriptstyle A}} 	\def\mB{m_{\scriptscriptstyle B}}
\def\LO{\Lambda^0}	\def\SO{\Sigma^0}	\def\SM{\Sigma^-}
\def\SP{\Sigma^+}	\def\SC{\Sigma^\pm}	\def\XM{\Xi^-}
\def\Kl3{K_{\ell3}}
\def\ftnt{\,\footnote}
\def\NoCol{\multicolumn{1}{c}{}}
\makeatletter
\newsavebox{\@bxnt}
\newlength{\@bxntwd}
\def\boxnotes#1{\sbox{\@bxnt}{#1} \settowidth{\@bxntwd}{\usebox{\@bxnt}}
  \begin{minipage}{\@bxntwd} \let\footnoterule=\relax #1 \end{minipage}}
\makeatother
\begin{document}
\title{\bf SU(3) Breaking Effects in\\
           Hyperon Semi-Leptonic Decays\\
           and the extraction of {\boldmath$F$} and {\boldmath$D$}\\[6pt]
 \normalsize\rm (submitted to \PL)}

\author{Philip G. Ratcliffe\\
 \small\it Dip.\ di Fisica, Univ.\ di Milano\\[-6pt]
 \small\it via G. Celoria 16, 20133 Milano, Italy\\
 \small\it E-mail: PGR@VAXMI.MI.INFN.IT}

\date{August 1995}

\maketitle

\begin{abstract}
The analysis of hyperon semi-leptonic decay data is re-examined in the light of
a recent suggestion that SU(3) symmetry breaking effects may be taken into
account by applying a correction to the $F/D$ ratio obtained via \naive\ linear
extrapolation in the hyperon masses. Comparison is made with the physically
better motivated approach of applying so-called centre-of-mass corrections.
This study (including all available data) reveals certain short-comings of the
former of the above treatments, highlights some interesting aspects of this
type of analysis and attempts to pinpoint the decay data that might usefully be
improved. A tantalising result of the SU(3) breaking analysis performed here is
that the magnitude of recoil correction required by the data corresponds
closely to that required for the standard explanation of the reduction of $\gA$
from its SU(6) value of 5/3. We also comment on other recent suggestions for
taking into account the effects of SU(3) breaking. Finally a few remarks are
made on the relevance for predicting the flavour non-singlet contribution to
the proton $g_1$ and the Ellis-Jaffe sum rule.
\\[6pt]
PACS: 13.30.Ce, 13.88.+e, 11.30.Hv, 13.60.Hb
\\[6pt]
hep-ph/9509237
\end{abstract}

\newpage
\section{Introduction}

In the wake of the EMC measurement~\cite{EMC88} of the spin-dependent proton
structure function $g_1^p(x)$, much
attention~\cite{LIP88,CLO89,RAT90,ROO90,DZI91,CLO93} has been
focussed on the $F/D$ ratio extracted from hyperon semi-leptonic decay (HSD)
data and used in predictions of the related Ellis-Jaffe sum
rule~\cite{ELL74}. In recent years the precision of experimental HSD data has
improved considerably~\cite{BOU82,HSU88,DWO90,PDG94} with some parameters and
rates now being measured to within an accuracy of a very few percent. Indeed,
such is the present accuracy that an approach for applying corrections due to
the breaking of SU(3) is now utterly indispensable. On the other hand, it has
in the past even been suggested that the description in terms of the usual $F$
and $D$ parameters should be abandoned altogether~\cite{LIP88}.

With regard to the analysis of the above-mentioned EMC and more recent SMC and
SLAC measurements of the nucleon spin structure functions~\cite{SMC94,SLA95},
a sizable shift in the $F/D$ ratio would remove the necessity for invoking a
large strange-quark spin component of the proton~\cite{EHR94} (referred to
here as ES). The size of the shift from the \naive\ SU(3)-based value depends
strongly upon the framework used to describe the violating effects. It is
important therefore to study the data with an eye to the sensitivity of the
$F/D$ ratio to the assumptions made as to the effects of SU(3) breaking, which,
in the past, have always been found to be at most of order 10\%, as might be
expected {\it a priori\/}.

In this letter we shall examine the present data and, in particular, attempt to
compare the various proposals for SU(3) breaking models and their effect on the
interpretation of the data. In the following section we present the data used
and, very briefly, the situation as regards the recurrent problem of
discrepancies in the life-time and angular correlation measurements in neutron
$\beta$-decay, and in section~\ref{sec:SU3} we present the results of an SU(3)
symmetric analysis as a reference point. In section~\ref{sec:brk} we examine
and compare the effects of various possible SU(3)-breaking scenarios. Thus,
after establishing the nature of the problem, in section~\ref{sec:recent} we
then examine recent proposals for dealing with SU(3) breaking. Finally, we
present conclusions and some indications of the relevance to the proton-spin
problem together with suggestions for further measurements aimed at better
understanding hyperon semi-leptonic decays.

\section{Hyperon Semi-Leptonic Decay Data}

The HSD data considered here are shown in table~\ref{tab:data}, which
represents the useful knowledge presently available.
\begin{table}[hbt]
\caption{The hyperon semi-leptonic data used in this
analysis~\protect\cite{PDG94}, $\gA/\gV$ is as extracted from the angular
correlations in the electron decay mode.}
\centering
\boxnotes{$\begin{array}
 {|l@{\,\to\,}l|l@{\,\pm\,}l|l@{\,\pm\,}l|l@{\,\pm\,}l|l|} \hline
 \multicolumn{1}{|c}{} & &
 \multicolumn{4}{c|}{\mbox{Rate ($10^6\,$s$^{-1}$)}} & \NoCol & & \\
 \cline{3-6}
 \multicolumn{2}{|c|}{\raisebox{2ex}[0pt][0pt]{Decay}}         &
 \multicolumn{2}{ c|}{\ell=e}                                  &
 \multicolumn{2}{ c|}{\ell=\mu}                                &
 \multicolumn{2}{ c|}{\raisebox{2ex}[0pt][0pt]{$\gA/\gV$}}     &
 \multicolumn{1}{ c|}{\raisebox{2ex}[0pt][0pt]{$\gA/\gV$}}
 \\ \hline  n     & p\,\ell\bar\nu & 1.1274  & 0.0025
\ftnt{Rate in $10^{-3}\,$s$^{-1}$.}
                                                          &
 \NoCol &         & \Z1.2599       & 0.0025
\ftnt{Taken from ref.~\protect\cite{TOW95a}.}
                                             & F+D
 \\ \hline \LO    & p\,\ell\bar\nu & 3.161   & 0.058      &
 0.60   & 0.13    & \Z0.718        & 0.015   & F+D/3
 \\ \hline \SM    & n\,\ell\bar\nu & 6.88    & 0.23       &
 3.04   & 0.27    & -0.340         & 0.017   & F-D
 \\ \hline \SM    & \LO\ell\bar\nu & 0.387   & 0.018      &
 \NoCol &         & \NoCol         &         & -\sqrt{\frac23}\,D
\ftnt{$\gV=0$, absolute expression for $\gA$ given.}
 \\ \hline \SP    & \LO\ell    \nu & 0.250   & 0.063      &
 \NoCol &         & \NoCol         &         & -\sqrt{\frac23}\,D\,^c
 \\ \hline \XM    & \LO\ell\bar\nu & 3.35    & 0.37
\ftnt{Scale factor 2 included, following the PDG practice for discrepant data.}
                                                          &
 2.1    & 2.1
\ftnt{Not used in fits.}
                  & \Z0.25         & 0.05    & F-D/3
 \\ \hline \XM    & \SO\ell\bar\nu & 0.53    & 0.10       &
 \NoCol &         & \NoCol         &         & F+D
 \\ \hline
\end{array}$}
\label{tab:data}
\end{table}
A first important point is the widely discussed discrepancy between the neutron
lifetime and the value of $\gA/\gV$ extracted from $\beta$-decay angular
correlations~\cite{TOW95a}. In recent years both quantities have been subject
to shifts and their experimental precision has significantly improved. The
present value of the neutron lifetime is $887.0\pm2.0\,$s and $\gA/\gV$ (from
angular correlations) is 1.2599(25)~\cite{TOW95a}, \ie, both are known
independently to approximately 0.2\%. Using the $ft$ values from the eight
super-allowed nuclear $\beta$-decay Fermi transitions, the relevant
Cabibbo-Kobayashi-Maskawa (CKM) matrix element is
$V_{ud}=0.9740(5)$~\cite{TOW95b}; to be compared with the values:
$V_{ud}=0.9795(20)$, from the neutron lifetime and $\gA/\gV$, and
$V_{ud}=0.9758(4)$, from the so-called $\Kl3$ decays
($V_{us}=0.2188(16)$~\cite{GAR92}).

Thus, there is no reason to assume any of the measurements to be more or less
reliable, especially in view of the fact that the CKM unitarity violation for
the two cases is of opposite sign. Moreover, the displacements from the central
values are $<0.2\%$, which nevertheless represents a much greater accuracy than
is presently necessary for HSD analysis, or indeed for comparison with the
Bjorken sum rule. Despite the interest in such a problem, it is beyond the
scope of the present paper and it is reasonable for the purpose of this
analysis to ignore the discrepancy; naturally though, its effects on the
$\chi^2$ of fits obtained will be taken into account.

\section{SU(3) Symmetric Analysis} \label{sec:SU3}

In order to have a clear idea of the problem within a phenomenological
framework, let us make a first attempt at globally fitting the HSD data. Apart
from a separation into lifetime and angular-correlation data, the distinction
can also be made between strangeness conserving and changing decays. In table
\ref{tab:su3fita} we present the results of a series of three-parameter ($F$,
$D$ and $V_{ud}$) fits to different sub-classes of the HSD data alone; no value
for $V_{ud}$ is imposed externally except for the final row, where we include
the mean value obtained from the combined nuclear $ft$ analysis and $\Kl3$
decays just described. We shall also impose the constraint
$V_{ud}^2+V_{us}^2=1$ and thus neglect $V_{ub}$
($V_{ub}=0.0032\pm0.0009~\cite{PDG94}$).
\begin{table}[hbt]
\caption{An SU(3) symmetric fit to the data.}
\centering
\boxnotes{$\begin{array}
 {|c|l@{\,\pm\,}l|l@{\,\pm\,}l|l@{\,\pm\,}l|c|c|} \hline
 & \multicolumn{6}{c|}{\mbox{Parameters}} & &        \\
 \cline{2-7}
 \multicolumn{1}{|c|}{\raisebox{2ex}[0pt][0pt]{Data}}          &
 \multicolumn{2}{ c|}{V_{ud}}                                  &
 \multicolumn{2}{ c|}{F}                                       &
 \multicolumn{2}{ c|}{D}                                       &
 \multicolumn{1}{ c|}{\raisebox{2ex}[0pt][0pt]{$\chi^2$/DoF}}  &
 \multicolumn{1}{ c|}{\raisebox{2ex}[0pt][0pt]{$F/D$}}         \\ \hline
 \mbox{Rates}      & 0.9749 & 0.0004 & 0.469 & 0.008 & 0.797 & 0.008
                   & 3.8    & 0.589 \\ \hline
 \gA/\gV           & \multicolumn{2}{c|}{-
\rlap{\ftnt{Undetermined.}}
                                          }
                                     & 0.460 & 0.008 & 0.800 & 0.008
                   & 0.8    & 0.576 \\ \hline
 \Y\Delta{S}\Y=0   & 0.9795 & 0.0020 & 0.528 & 0.017 & 0.732 & 0.017
                   &  -
\rlap{\ftnt{Zero degrees of freedom.}}
                            & 0.721 \\ \hline
 |\Delta{S}|=1     & 0.9742 & 0.0006 & 0.448 & 0.009 & 0.791 & 0.017
                   & 0.8    & 0.567 \\ \hline
 \mbox{All}        & 0.9750 & 0.0004 & 0.465 & 0.006 & 0.799 & 0.006
                   & 3.0    & 0.582 \\ \hline
 \mbox{All}+V_{ud}
\ftnt{$V_{ud}$ from nuclear $ft$ and $\Kl3$ analysis.}
                   & 0.9751 & 0.0002 & 0.465 & 0.006 & 0.799 & 0.006
                   & 2.7    & 0.583 \\ \hline
\end{array}$}
\label{tab:su3fita}
\end{table}

Two interesting points emerge from this fit: first of all, the value of
$V_{ud}$ obtained solely from the hyperon data is consistent with that coming
from the nuclear $ft$ analysis and $\Kl3$ data. Second, it should be pointed
out that more than half the total $\chi^2$ (neglecting the discrepant neutron
data contribution) comes from the $\SC\to\LO\ell\nu$ data alone (where only the
rates are accessible). This is interesting when considered together with the
fact that these decays are unique in having no vector contribution and
therefore, as we shall see later, have the rates most affected by the recoil
corrections. Moreover, when fit together only with the other $|\Delta{S}|=0$
decays, they provide a value for $F/D$ close to the SU(6) prediction. This
suggests that the problem may arise owing to the bulk of lesser affected data
forcing a particularly poor value onto the one very sensitive point and not (as
has been suggested~\cite{ROO90}) that this experimental rate may be wrong; we
shall return to this later.

Notice finally that, within the errors, the angular correlation data alone is
well described by an SU(3) symmetric fit. Thus, one sees the futility of trying
to extract any information on SU(3) breaking using these data alone and, as
discussed later, the large errors arising in the ES analysis are partially
explained.

Thus, to eliminate the effect of the neutron problem on the global $\chi^2$ we
first extract a mean value for $V_{ud}$ from the nuclear $ft$ and $\Kl3$ data;
using this value, we make a combined fit to the neutron rate and $\gA/\gV$.
Then, in the absence of any indication as to where the problem may lie, we
multiply the errors of the neutron lifetime, $\gA$ and mean $V_{ud}$ values by
the $\sqrt{\chi^2}$ so obtained, and use these in all the following fits:
\begin{eqnarray}
 \mbox{Rate}(n\to p\ell\bar\nu) &=& (1.1274 \pm 0.0055) \times 10^{-3}s^{-1} \\
 \gA/\gV                        &=& \hphantom{(} 1.2599 \pm 0.0055           \\
 V_{ud}                         &=& \hphantom{(} 0.9752 \pm 0.0007;
\end{eqnarray}
in table~\ref{tab:su3fitb} we display the SU(3) symmetric fit results using
these values.
\begin{table}[hbt]
\caption{An SU(3) symmetric fit to the modified data including the external
$V_{ud}$ from nuclear $ft$ and $\Kl3$ analysis (see text for details).}
\centering
\boxnotes{$\begin{array}
 {|c|l@{\,\pm\,}l|l@{\,\pm\,}l|l@{\,\pm\,}l|c|c|} \hline
 & \multicolumn{6}{c|}{\mbox{Parameters}} & &        \\
 \cline{2-7}
 \multicolumn{1}{|c|}{\raisebox{2ex}[0pt][0pt]{Data
                                                   }}          &
 \multicolumn{2}{ c|}{V_{ud}}                                  &
 \multicolumn{2}{ c|}{F}                                       &
 \multicolumn{2}{ c|}{D}                                       &
 \multicolumn{1}{ c|}{\raisebox{2ex}[0pt][0pt]{$\chi^2$/DoF}}  &
 \multicolumn{1}{ c|}{\raisebox{2ex}[0pt][0pt]{$F/D$}}         \\ \hline
 \mbox{Rates}      & 0.9749 & 0.0003 & 0.469 & 0.008 & 0.796 & 0.009
                   & 3.2    & 0.589 \\ \hline
 \gA/\gV           & 0.9752 & 0.0007 & 0.460 & 0.008 & 0.799 & 0.009
                   & 0.8    & 0.576 \\ \hline
 \Y\Delta{S}\Y=0   & 0.9753 & 0.0007 & 0.529 & 0.017 & 0.735 & 0.017
                   & 0.5    & 0.719 \\ \hline
 |\Delta{S}|=1     & 0.9747 & 0.0005 & 0.452 & 0.009 & 0.799 & 0.015
                   & 0.8    & 0.566 \\ \hline
 \mbox{All}        & 0.9749 & 0.0003 & 0.465 & 0.006 & 0.798 & 0.006
                   & 2.3    & 0.582 \\ \hline
\end{array}$}
\label{tab:su3fitb}
\end{table}
It can be seen that, having taken account of the neutron discrepancy, there is
still a problem (stemming, as before, from the $\SC\to\LO\ell\nu$ decay rate).
Also worthy of remark is the fact that the $|\Delta{S}|=0$ and 1 data are
separately well fit, suggestive of some correlated effect; we shall return to
this later.

\section{SU(3) breaking corrections} \label{sec:brk}

The next step is to consider possible corrections to these processes. It has
long been known that a realistic explanation of the renormalisation of the
neutron $\beta$-decay $\gA$ can be provided in terms of relativistic
corrections~\cite{BOG68}. Such an approach has already been applied with
success to the HSD data~\cite{DHK87,RAT90} and it has been noted that there
might even be evidence that this accounts for SU(6) breaking~\cite{CLO93}.

One of the earliest analyses of this type~\cite{DHK87} (referred to here as
DHK) also attempted to include the effects of wave-function mismatch between
the strange and $u,d$ quarks. However, the newer more precise data now strongly
suggest that, at least as calculated there, such an effect is not present.
On the other hand, since the data do seem to suggest some sort of correlated
dependence, we shall also examine the effect of introducing an \adhoc\
correction for the $|\Delta{S}|=1$ decays.

The DHK approach is to apply centre-of-mass (CoM) or recoil corrections to the
axial coupling constant for the process $A\to{B}\ell\nu$ according to the
following formula~\cite{DHK87}:
\begin{equation}
 \gA = \gA^{\mathrm{SU(3)}} \; \left\{ 1 - \frac{\vev{p^2}}{3\mA\mB} \,
 \left[ \frac14 + \frac{3\mB}{8\mA} + \frac{3\mA}{8\mB} \right] \right\},
\label{eq:CoM}
\end{equation}
with a similar correction to the vector piece, which is entirely negligible (in
accordance with the Ademollo-Gatto theorem~\cite{ADE64}). The mean squared
momentum, $\vev{p^2}$, is calculated by DHK using a bag model to be
$0.43\gev^2$. In their analysis DHK use the ratio of the correction to that for
the neutron (taken as a convenient reference value) since for the purposes of
such a fit the important quantity is precisely this ratio. Thus, we begin the
SU(3) breaking analysis using the DHK approach: the results are presented in
table~\ref{tab:CoM}.
\begin{table}[hbt]
\caption{An SU(3) breaking fit to the modified data including the external
$V_{ud}$; only the approximate (DHK) CoM correction is applied (see text for
details).}
\centering
\boxnotes{$\begin{array}
 {|c|l@{\,\pm\,}l|l@{\,\pm\,}l|l@{\,\pm\,}l|c|c|} \hline
 & \multicolumn{6}{c|}{\mbox{Parameters}} & &        \\
 \cline{2-7}
 \multicolumn{1}{|c|}{\raisebox{2ex}[0pt][0pt]{Data
                                                   }}          &
 \multicolumn{2}{ c|}{V_{ud}}                                  &
 \multicolumn{2}{ c|}{F}                                       &
 \multicolumn{2}{ c|}{D}                                       &
 \multicolumn{1}{ c|}{\raisebox{2ex}[0pt][0pt]{$\chi^2$/DoF}}  &
 \multicolumn{1}{ c|}{\raisebox{2ex}[0pt][0pt]{$F/D$}}         \\ \hline
 \Y\Delta{S}\Y=0   & 0.9753 & 0.0007 & 0.481 & 0.018 & 0.784 & 0.018
                   & 0.5    & 0.613 \\ \hline
 |\Delta{S}|=1     & 0.9747 & 0.0005 & 0.465 & 0.009 & 0.825 & 0.015
                   & 1.0    & 0.563 \\ \hline
 \mbox{All}        & 0.9744 & 0.0003 & 0.460 & 0.006 & 0.806 & 0.006
                   & 1.0    & 0.571 \\ \hline
\end{array}$}
\label{tab:CoM}
\end{table}

The improvement is quite dramatic and is immediately seen to be essentially due
to a sizable shift in the parameter values of the $\Delta{S}=0$ fit. The main
effect is to increase (by about 6\%) the value of $D$ obtained from the
$\SC\to\LO\ell\nu$ decay rates (where recall $\gV=0$) and correspondingly
reduce the value of $F$, thus bringing these decays into line with the rest
(where the effect is more modest and acts to increase both $F$ and $D$
simultaneously). The overall $\chi^2$ is good and no single data point stands
out as particularly poorly fit: the worst is $\gA/\gV$ for
$\XM\to\LO{e}\bar\nu$, which contributes 2.1 to the total $\chi^2$.

We note in passing that inclusion of the strange-quark wave-function mismatch
correction, \`a la DHK, worsens the fits: without the CoM correction we obtain
$\chi^2=5.6$ and with $\chi^2=2.6$ (for all data). One other observation of
interest is that $\vev{p^2}=0.43$ actually corresponds to the best-fit value
for this parameter, thus increasing the confidence in such an approach.
Finally, note that the overall values of $F$ and $D$ have only been shifted by
less than 1\% and the ratio $F/D$ by less than 2\%.

Now we come to the possibility of a link with the SU(6) wave functions. The
form of the correction given in eq.~(\ref{eq:CoM}) is an $O(p^2)$ approximation
to the exact expression and while the shift applied to the $\gA$ above (after
dividing out the neutron correction) is always less than 8\%, the individual
corrections are much larger and the approximation is rather poor. The exact
form of the correction for $A\to{B}\ell\nu$ may be written as
\begin{equation}
 \def\epsil^#1_#2{\epsilon^#1_{\scriptscriptstyle#2}}
\begin{array}{rcl}
 \gV &=& \gV^{\mathrm{SU(3)}} \;
         \left[ \epsil^+_A \epsil^+_B + \epsil^-_A \epsil^-_B \right] \\
 \gA &=& \gA^{\mathrm{SU(3)}} \;
         \left[ \epsil^+_A \epsil^+_B - \frac13 \epsil^-_A \epsil^-_B \right],
\end{array}
\end{equation}
where $\epsilon^\pm_i=\sqrt{(E_i\pm m_i)/2E_i}$, with $E_i=\sqrt{m_i^2+p^2}$.
Again the correction to the vector coupling is never more than 0.2\% and is
thus negligible. Requiring that this reduce the SU(6) value of the neutron
$\gA$ from 5/3 to $\sim5/4$, fixes $p^2=1.3\gev^2$ (which, \aposteriori,
demands use of the exact form of the expression). Choosing this value of $p^2$
and applying the exact formula we obtain the results shown in
table~\ref{tab:exact}.
\begin{table}[hbt]
\caption{An SU(3) breaking fit to the modified data including the external
$V_{ud}$; the exact CoM correction is applied (see text for details).}
\centering
\boxnotes{$\begin{array}
 {|c|l@{\,\pm\,}l|l@{\,\pm\,}l|l@{\,\pm\,}l|c|c|} \hline
 & \multicolumn{6}{c|}{\mbox{Parameters}} & &        \\
 \cline{2-7}
 \multicolumn{1}{|c|}{\raisebox{2ex}[0pt][0pt]{Data}}          &
 \multicolumn{2}{ c|}{V_{ud}}                                  &
 \multicolumn{2}{ c|}{F}                                       &
 \multicolumn{2}{ c|}{D}                                       &
 \multicolumn{1}{ c|}{\raisebox{2ex}[0pt][0pt]{$\chi^2$/DoF}}  &
 \multicolumn{1}{ c|}{\raisebox{2ex}[0pt][0pt]{$F/D$}}         \\ \hline
 \Y\Delta{S}\Y=0   & 0.9753 & 0.0007 & 0.480 & 0.018 & 0.785 & 0.018
                   & 0.5    & 0.611 \\ \hline
 |\Delta{S}|=1     & 0.9747 & 0.0005 & 0.464 & 0.009 & 0.825 & 0.015
                   & 1.0    & 0.563 \\ \hline
 \mbox{All}        & 0.9744 & 0.0004 & 0.460 & 0.006 & 0.806 & 0.006
                   & 1.0    & 0.570 \\ \hline
\end{array}$}
\label{tab:exact}
\end{table}

The unexpected result is a fit almost identical to the original approximate CoM
correction with, however, the difference that here the choice of the parameter,
$p^2$, was guided by the desire to explain the shift in the neutron $\gA/\gV$
from its SU(6) value; again it turns out to be very close to the best-fit
value. Having said that, it is obvious that, although in this way we have
``restored'' the SU(6) picture for $F+D$, the individual SU(6) values of $F$
and $D$ are not recovered. Nonetheless, the $\Delta{S}=0$ decays do return a
value still close to $F/D=2/3$.

To close this section, we remark on the possibility of including the type
of wave-function mismatch correction mentioned above. In their bag-model
calculation DHK arrived at an enhancement of the axial coupling by 8\% due to
this effect (while the vector coupling was reduced by 1.3\%). In the analysis
performed here we have consistently found a preference for a small enhancement
of both the axial and vector couplings, by about 2\%. However, within errors,
the results are also consistent with zero effect. In other words, if both
$V_{ud}$ and $V_{us}$ are allowed to float a net tendency for over-saturation
of CKM unitarity is observed (independently of whether or not external
constraints on $V_{ud}$ are imposed). In all cases the ratio, $F/D$, is
affected by at most 0.5\%. We note also that this explains the rather large
value for $V_{ud}$ found in an earlier such analysis~\cite{RAT90}, which can
thus be accounted for by a small renormalisation of the strangeness-changing
couplings.

\section{Recent approaches to SU(3) breaking} \label{sec:recent}

Inspired by the observation that the $F/D$ ratios extracted from the
angular-correlation measurements display an approximately linear variation with
the mass difference of the relevant hyperons with respect to the proton and
neutron, Ehrnsperger and Sch\"afer~\cite{EHR94} have attempted an
SU(3)-breaking analysis and extraction of the $F$ and $D$ parameters. The idea
is simply that the neutron $\beta$-decay $\gA/\gV$ provides the sum, $F+D$, and
the remaining three known data values are used to make a one-parameter fit to
an \adhoc\ correction:
\begin{equation}
 F/D = (F/D)^{\mathrm{SU(3)}} \;
       \left[ 1+a\frac{(\mA+\mB)-(m_n+m_p)}{(\mA+\mB)+(m_n+m_p)} \right],
\end{equation}
where $a$ is found to be $\sim2.7$ and the limiting value of $F/D$ (valid for
the nucleons) is $0.49\pm0.08$.

There are several criticisms to be levelled at such an approach, both of a
formal theoretical nature and of a more practical type. First of all, let us
recall that $F$ and $D$ are simply the antisymmetric and symmetric (in flavour
indices) reduced matrix elements for charged-current baryon-baryon transitions.
Thus, it is very hard to see why they should be renormalised in such a way as
to miraculously maintain their sum constant while changing their ratio.
Moreover, such a solution would imply that the decay $\XM\to\SO\ell\bar\nu$
should have $\gA/\gV$ identical to that of the neutron, despite the enormous
mass difference. Indeed, the breaking pattern so-predicted appears to be
entirely random when viewed from the point-of-view of the various $\gA/\gV$.
Put simply, there is an implicit, arbitrary and unexplained assumption in the
ES approach: namely, that the particular combination $F+D$ is protected against
SU(3) breaking.

On the practical side, the first and obvious objection is the neglect of the
decay-rate data; we have repeated the ES analysis including all the data and
find that the value of the breaking parameter, $a$, rises to $\sim7$. While the
overall $\chi^2$ is still admittedly very good, the assumption of linearity is
now severely undermined and for certain decays ($\XM\to\LO\ell\bar\nu$) the
effective renormalisation of $\gA/\gV$ is over 500\%. The second objection has
to do with form chosen: it is a) linear in the mass breaking and b) neglects
the mass difference between initial and final state (\cf. the CoM correction).
In the light of our findings and, \eg, the large $\SC{-}n$ mass difference,
neither assumption is justifiable or even plausible.

Moreover, as noted above, the angular-correlation data alone are well
described even without SU(3) breaking. Thus, given the small lever-arm they
offer (the only two precise data points lie very close in terms of the
mass-breaking variable of ES), it is not surprising that this approach results
in a very different value of $F/D$ with also very large errors. Note that,
within the quoted errors, it does not significantly disagree with any of the
other analyses

Let us now address the various approaches in which the validity of SU(3), and
its use, is seriously questioned~\cite{DZI91,LIP88,FRA84,LIP94,LIC95}. One of
the main objections raised is the difficulty in explaining the baryon magnetic
moments. While in no way wishing to embark on a discussion that would take us
beyond the scope of this paper, we would counter such objections with two
observations: firstly, the magnetic moments are exquisitely linked to the
masses of the quarks, which (by definition) are unknown, while the axial
couplings should not suffer such a complication. Second, a not entirely
unsatisfactory picture of the magnetic moments can be obtained: the application
of a similar correction to that used here allows a description that is good to
within 0.07 nuclear magnetons over the whole octet (and also the $\Omega^-$
decuplet resonance)~\cite{RAT95}. It should always be borne in mind that a
\naive\ two-parameter fit to the nucleon magnetic moments already reveals one
of the important problems that makes this such a delicate subject:
$-2\mu_d/\mu_u=1.05$, indicating a 5\% violation of isospin symmetry.

On the other hand, various of the proposals (see, \eg, \cite{LIP94,LIC95}) do
not imply a failure of the SU(3) description of the hyperon semi-leptonic
decays but instead call into question the direct use of $F$ and $D$ in
separating the quark spin contributions. However, the main thrust of the
discussion presented here has been to demonstrate the applicability of SU(3)
(and its known violations) to the extraction of the $F$ and $D$ parameters,
without attempting to make any connection to the individual quark spins.

\section{Conclusions} \label{sec:conc}

Before closing, let us try and identify those decays that could (with more
precise measurement) throw useful light on this problem.
\begin{enumerate}
\item
The $\SC\to\LO\ell\nu$ decays are very significant. Firstly, they represent the
major problem for an SU(3) symmetric fit and any significant increase in the
measured rate for $\SM\to\LO{e}\bar\nu$ would greatly alleviate the situation.
Second, as the only $\Delta{S}=0$ decays besides the neutron, they are valuable
in helping to fix a reference point from which to judge the importance of
corrections in the $|\Delta{S}|=1$ case.
\item
The angular correlations in $\SM\to{n}e\nu$ and $\LO\to{p}e\nu$ could give
vital information on the importance of second-class currents, which in turn can
quite dramatically affect the values of $\gA$ extracted and again seem to
reduce the symmetry breaking necessary.
\item
The $\XM\to\SO{e}\nu$ rates and angular correlations, having the largest
corrections, are also more sensitive, in principle, to any breaking. Moreover,
the fact that $\gA=F+D$ for this process, identically to neutron $\beta$-decay,
makes it uniquely interesting.
\item
As a general observation, the angular correlations do not necessitate the
inclusion of the CKM matrix elements in the fit and thus suffer less the
ambiguity due to the, albeit small, discrepancy there.
\end{enumerate}

The one blemish on the analysis presented here is that, despite motivating the
corrections as being over an SU(6) symmetric ``background'', the values of $F$
and $D$ that finally emerge are still somewhat shifted. In mitigation of this
failure we would remind the reader that the strangeness conserving data alone
lie very close to the SU(6) picture. Moreover, inclusion of a possible
correction for the strange-quark wave function does tend to raise the fit value
of $F/D$. It was noticed in ref.~\cite{CLO93} that the experimental data on the
$\SM\to{n}e\bar\nu$~\cite{HSU88} decay also indicate a possible non-negligible
second-class current contribution. Indeed, the data preferred a sizable $g_2$
and thus a much reduced value for $\gA$ ($\sim0.2$). If this were the case
then, firstly, the data would approach much more closely the SU(6) expectations
(on inclusion of the corrections discussed here) and, secondly, the question
arises of the relevance of such currents in other decays, where the
experimental analysis has typically assumed them to be zero.

As for the implications in polarised DIS, the situation there is somewhat
clouded by the inherent ambiguities (essential in PQCD and particularly due to
the r\^ole of the anomaly) in defining separate quark-spin densities. The
values of the parameters obtained here are very much in line with those
generally used in the literature and so would not significantly alter the
conclusions. Thus, the standard PQCD resolution of the Ellis-Jaffe sum rule
discrepancy still demands some form of non-negligible strange quark
polarisation.

We have hopefully convinced the reader that the use of SU(3) symmetry with well
motivated corrections for its violation allows a very satisfactory description
of the hyperon semi-leptonic decays and leaves little room for any further
SU(3) breaking contributions. In support of the statement of validity we would
remind the reader of the remarkable success of the fit motivated by the
renormalisation of the neutron $\gA$ from 5/3 to 5/4: in this approach {\it no
new free parameter\/} was introduced and the initial $\chi^2$/DoF of over 2 was
reduced to 1. As far as other contributions or forms of correction are
concerned, our analysis has shown that there is little space for radically
different approaches when compared with the complete set of up-to-date
experimental data.

\newpage

\end{document}